\documentclass{PoS}
\usepackage{cite,amsmath}
\newcommand{\calO}{{\cal O}}

\title{Off-shell Higgs signal and total width determination at the LHC}

\ShortTitle{Off-shell Higgs signal and total width determination at the LHC}

\author{\speaker{Nikolas Kauer}\\
        Department of Physics, Royal Holloway, University of London, Egham TW20 0EX, UK\\
        E-mail: \email{n.kauer@rhul.ac.uk}}

\abstract{%
A substantial off-shell Higgs boson signal in the 
gluon fusion and vector boson fusion $H\to ZZ$ and $H\to WW$ channels
at the Large Hadron Collider (LHC) facilitates a novel, 
complementary approach to constraining the total Higgs width $\Gamma_H$. 
With LHC Run 1 data, experimental analyses by CMS and ATLAS find 
$\Gamma_H < 5.4\,\Gamma_H^\text{SM}$ and $\Gamma_H < [4.5,7.5]\,\Gamma_H^\text{SM}$ at 95\% confidence level, respectively, where $\Gamma_H^\text{SM}$ is the expected value in the Standard Model at the measured Higgs boson mass. I review the theoretical 
basis of the new approach and discuss its significance in comparison 
to other methods to bound and measure the Higgs width at the LHC and future 
colliders.
}

\FullConference{Frontiers of Fundamental Physics 14\\
		 15-18 July 2014\\
		 Aix Marseille University (AMU) Saint-Charles Campus, Marseille, France}

\begin{document}




\section{Introduction}

The fundamental particle predicted by the Standard Model (SM) Higgs mechanism 
\cite{HiggsTheory}, i.e.\ the Higgs boson, 
was discovered at the LHC in 2012 \cite{HiggsExperiment}.
A thorough examination has since taken place and its properties have been found to 
be in agreement with theoretical expectations. No compelling deviations from the 
SM have been discovered so far.  An important property of the Higgs 
boson is its total decay width, with a predicted SM value of 
$\Gamma_H^\text{SM}\approx 4$ MeV, which is more than two orders of magnitude
smaller than the experimental Higgs mass resolution at the LHC, which is of order 
$1$ GeV.  At the LHC, any direct Higgs width measurement via the resonance shape
is thus limited to an uncertainty of $\Delta\Gamma_H\sim 1$ GeV.\footnote{%
For instance, Ref.\ \cite{Chatrchyan:2013mxa} finds $\Gamma_H < 3.4$ GeV at 95\% confidence level (CL).}
Since the resonant (``on-peak'') Higgs cross section depends on $\Gamma_H$, 
the Higgs couplings and width cannot be determined independently 
at the LHC without relying on theoretical assumptions \cite{Duhrssen:2004cv,Dittmaier:2012vm}.  For instance, in models without triplett or higher $SU(2)$ representations an upper limit for the $HWW$ or $HZZ$ coupling exists and an upper bound for the 
Higgs width can be obtained that is of the order of the SM Higgs width 
\cite{Peskin:2012we,Dobrescu:2012td}.
Assuming no beyond-SM (BSM) Higgs decays, and suggestive Higgs coupling 
parameterisations, one can fit the Higgs width to data and finds similar agreement 
with $\Gamma_H^\text{SM}$ \cite{Barger:2012hv,Cheung:2013kla,Ellis:2013lra,Djouadi:2013qya,Bechtle:2014ewa}.
At a future $e^+e^-$ collider, a largely model-independent indirect determination
of the Higgs width will be possible, with a predicted accuracy of $5$--$10$\%
at the International Linear Collider \cite{Han:2013kya,Peskin:2013xra,Bechtle:2014ewa}.
A future muon collider could permit a direct Higgs width measurement via threshold 
scan with an estimated accuracy of $4$--$9$\% \cite{Han:2012rb}.
Two novel, complementary methods to constrain the Higgs width at the LHC 
are reviewed in Section \ref{main}.

\section{Off-shell Higgs signal enabled total width determination at the LHC\label{main}}

\begin{figure}
\begin{center}
\includegraphics[height=1.5cm, clip=true]{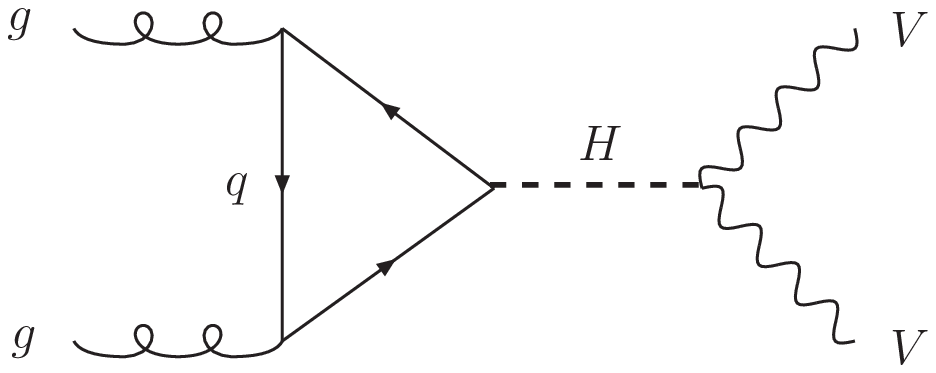}\hspace*{0.3cm}
\includegraphics[height=1.5cm, clip=true]{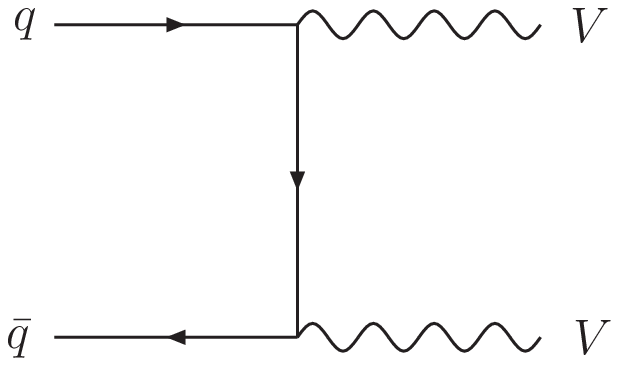}\hspace*{0.3cm}
\includegraphics[height=1.5cm, clip=true]{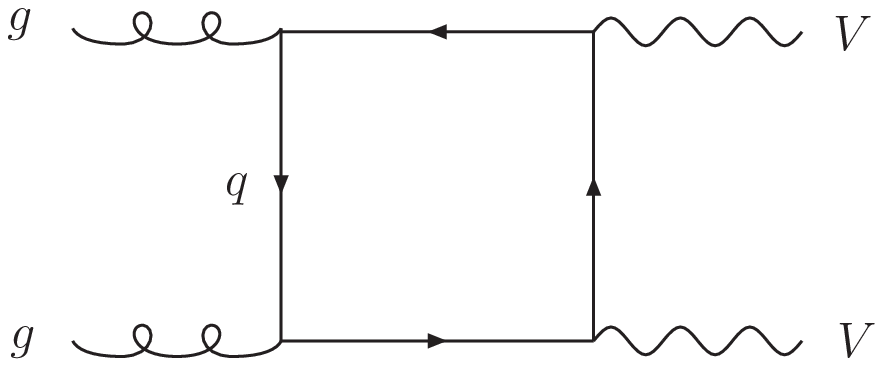}\\[0.8cm]
\includegraphics[height=5.cm, clip=true]{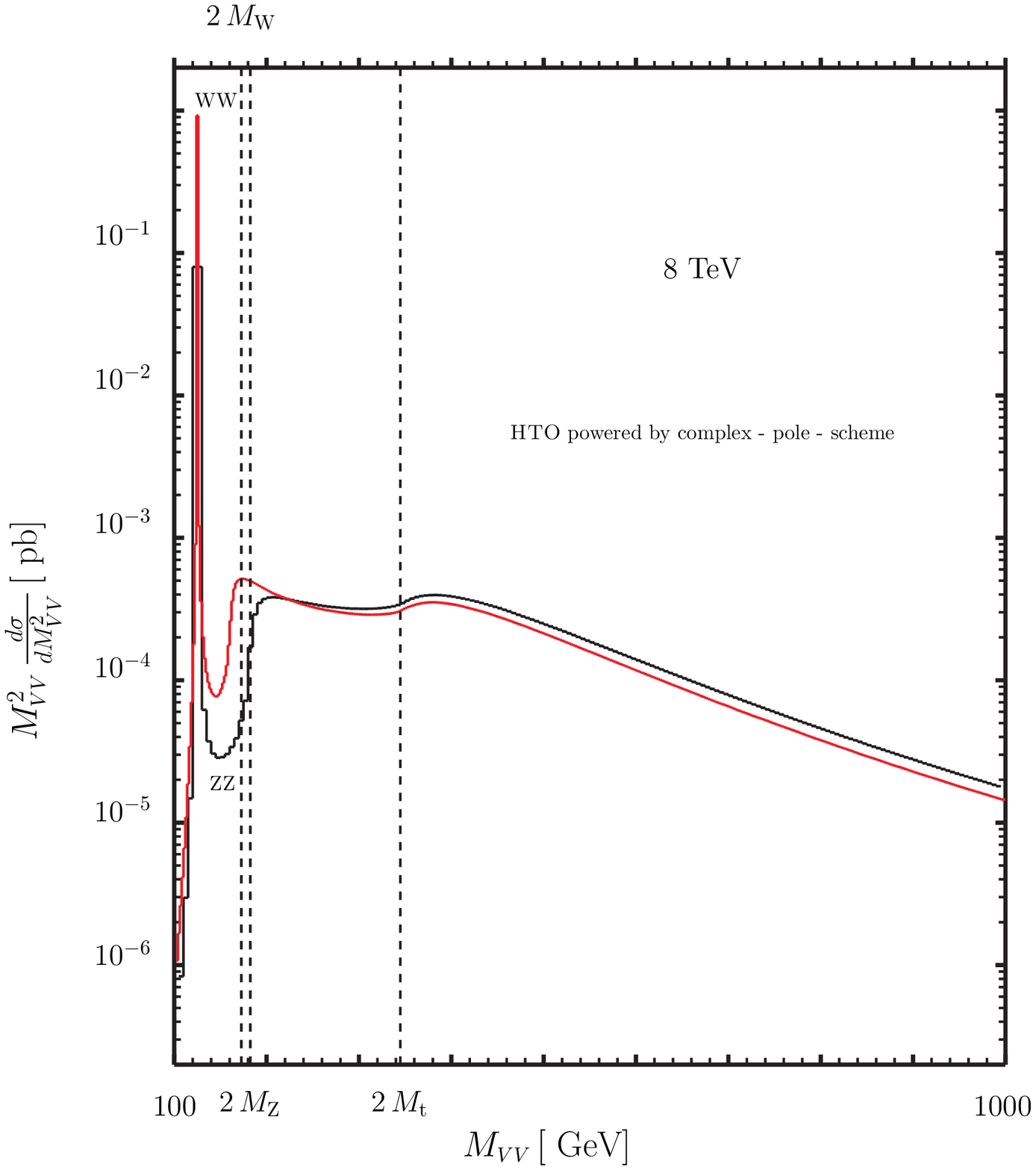}\hfil
\includegraphics[height=5.cm, clip=true]{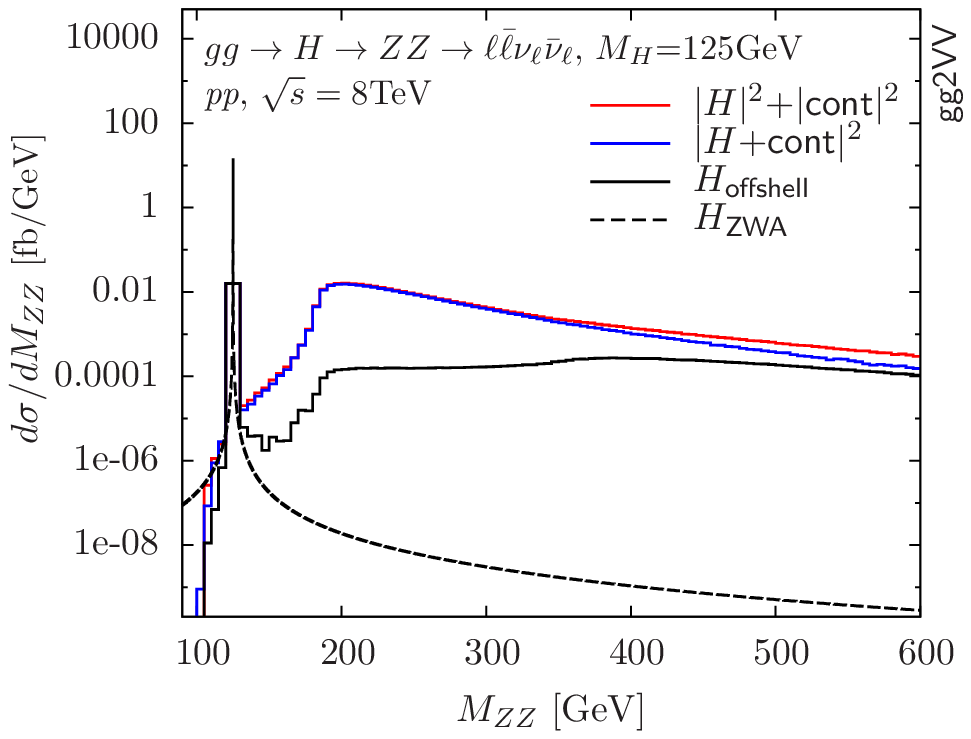}
\end{center}
\caption{Representative Feynman graphs for the $gg\to H\to VV$ signal process (left) and the $q\bar{q}$- (centre) and $gg$-initiated (right) continuum background processes at LO 
as well as $M_{VV}$ distributions that show the enhanced off-shell Higgs  signal and sizeable Higgs-continuum interference (from \cite{Kauer:2012hd}).}
\label{fig:tail}
\end{figure}

The existence of a substantial off-shell Higgs boson signal in the 
gluon fusion $H\to ZZ$ and $H\to WW$ channels
at the LHC was first pointed out in Ref. \cite{Kauer:2012hd}.\footnote{%
The significance of the off-shell $H\to VV$ signal 
at a linear collider is discussed in 
Ref.\ \cite{Liebler:2015aka}.}
In Fig. \ref{fig:tail}, representative graphs for the Higgs signal 
and continuum background processes are shown as well as $M_{VV}$ distributions 
that show the enhanced off-shell Higgs signal, which constitutes an 
$\calO(5$--$10\%)$ correction to inclusive $gg\to H\to VV$
production in narrow-width approximation (NWA).  With typical experimental LHC 
selection cuts this correction increases to $\calO(10$--$20\%)$.
Also shown in Fig.\ \ref{fig:tail} is the sizeable signal-background 
interference in the off-shell region, which facilitates unitarity at high 
energies and has been calculated in Refs.\ \cite{Glover:1988fe,Glover:1988rg,Binoth:2006mf,Campbell:2011cu,Kauer:2012hd,Bonvini:2013jha,Kauer:2013qba,Ellis:2014yca,Campanario:2012bh}.  Note that the interfering $gg\to VV$ continuum background at LO is 
formally part of the NNLO corrections to $pp\to VV$ \cite{Cascioli:2014yka,Gehrmann:2014fva}.

\begin{figure}
\begin{center}
\includegraphics[height=5.5cm, clip=true]{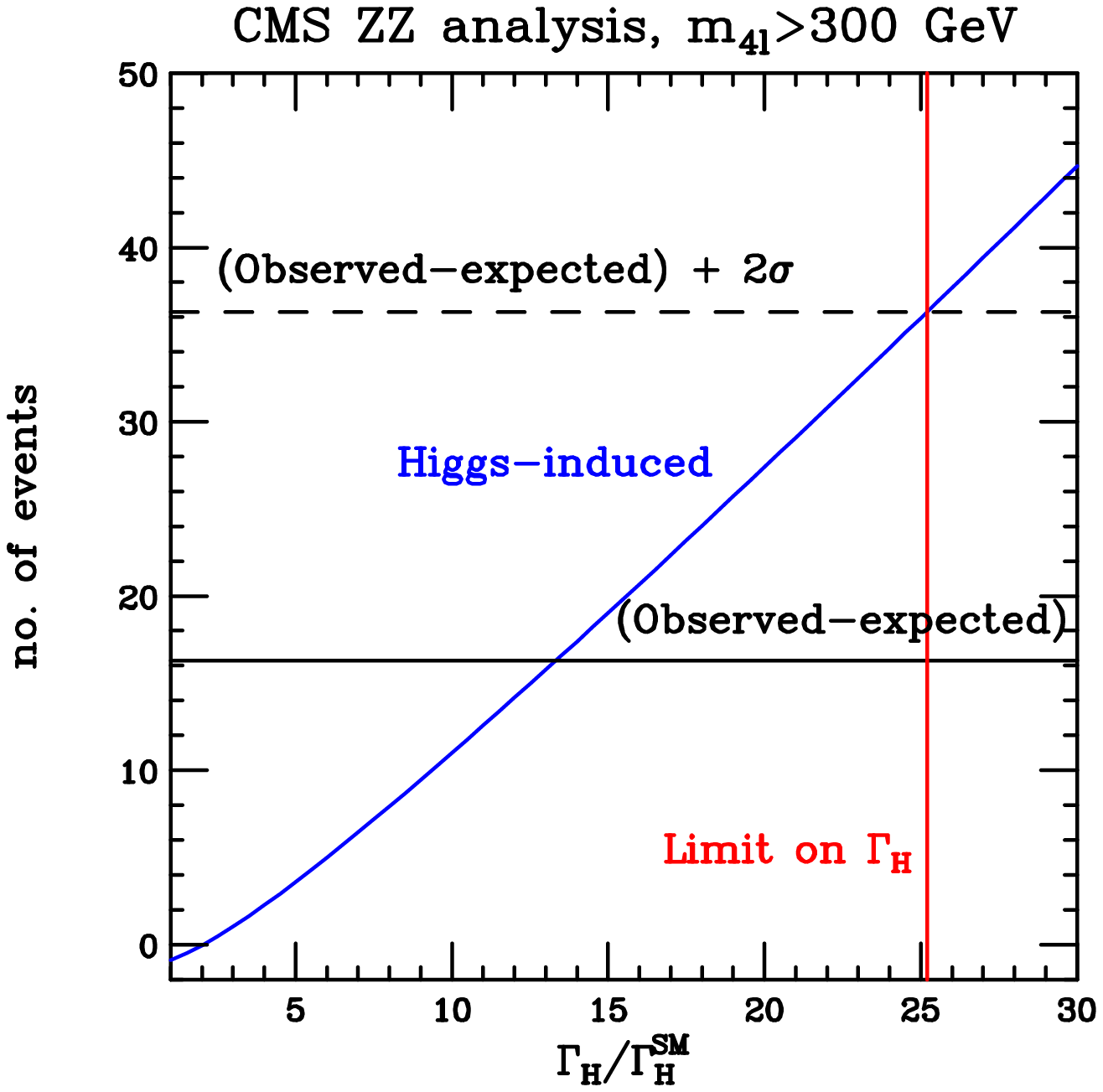}\hfil
\includegraphics[height=5.5cm, clip=true]{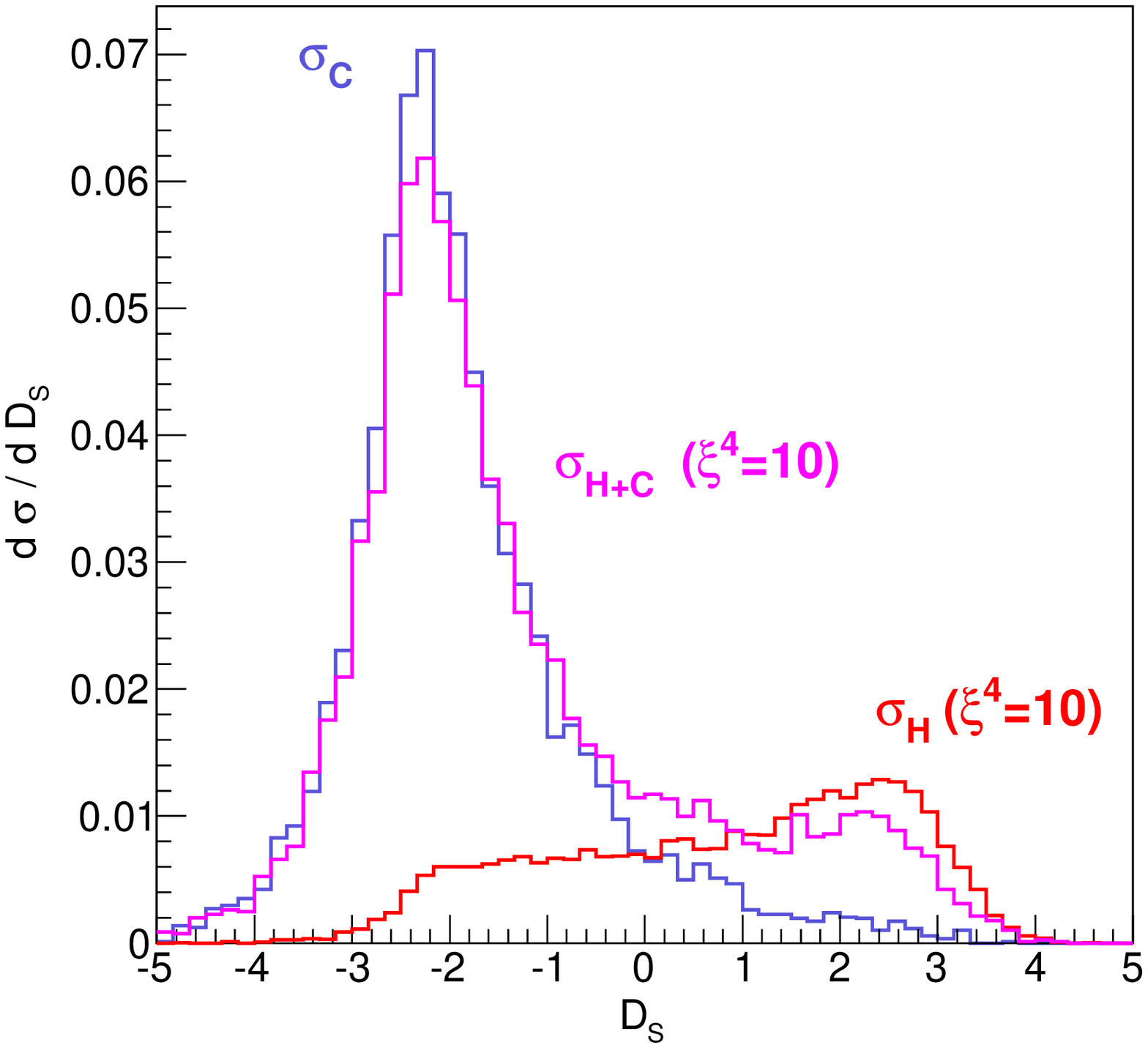}
\end{center}
\caption{Results of a detailed phenomenological study of the off-shell Higgs width constraint approach (from \cite{Campbell:2013una}).}
\label{fig:MCFM}
\end{figure}

\begin{figure}
\begin{center}
\includegraphics[height=5.5cm, clip=true]{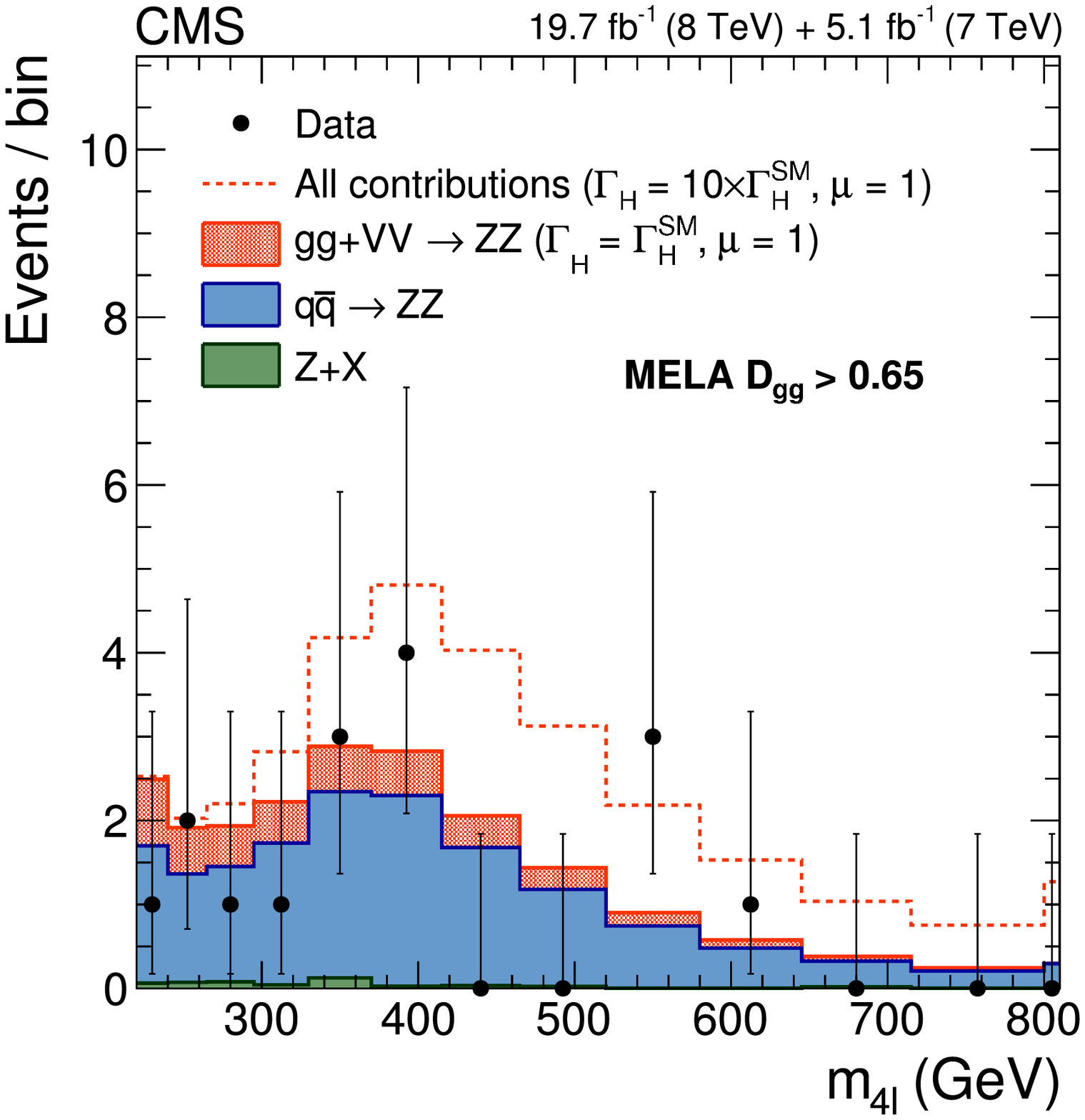}\hfil
\includegraphics[height=5.5cm, clip=true]{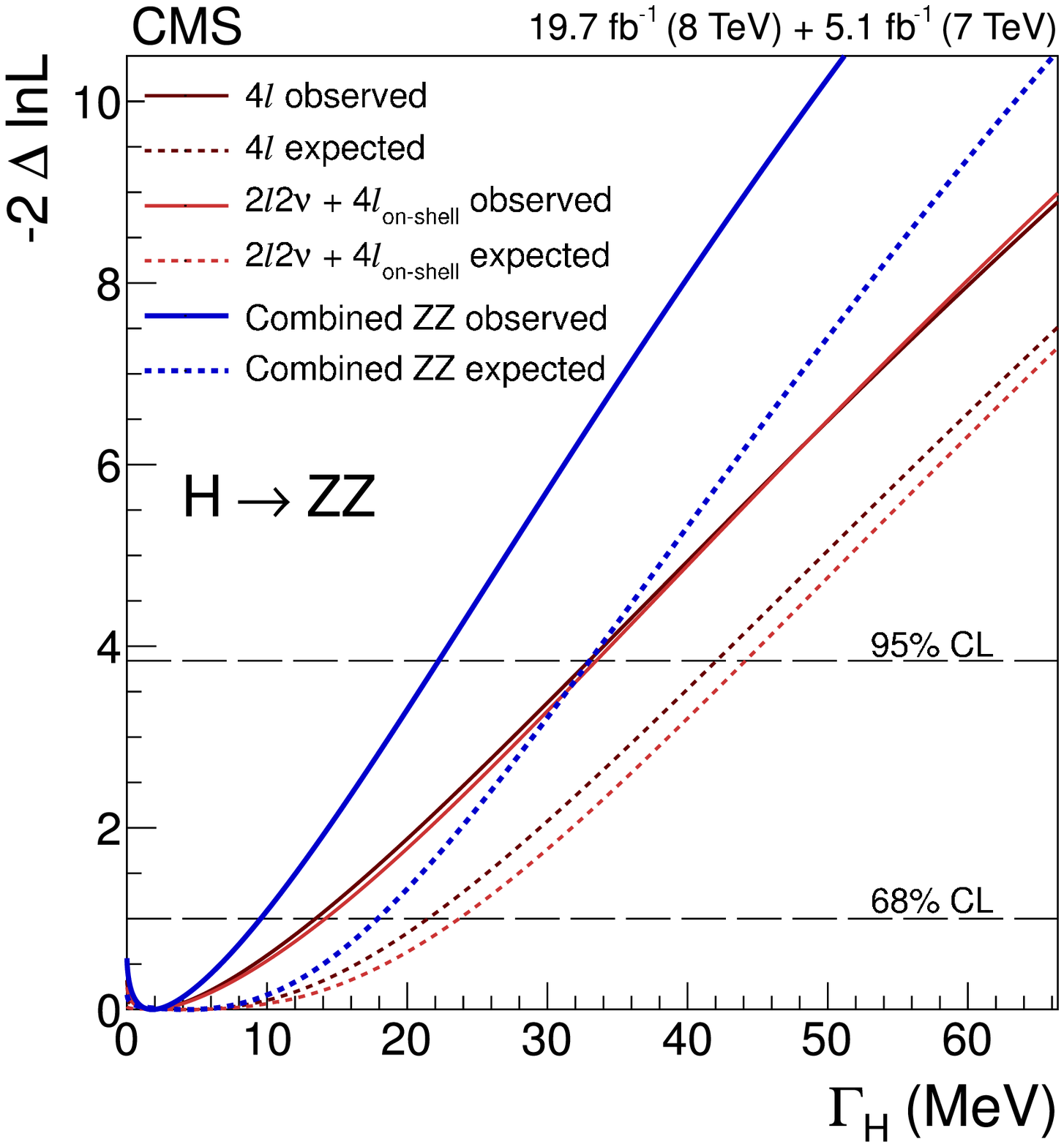}
\end{center}
\caption{CMS study of off-shell Higgs width constraint approach (from \cite{Khachatryan:2014iha}).}
\label{fig:CMS}
\end{figure}

\begin{figure}
\begin{center}
\includegraphics[height=4.8cm, clip=true]{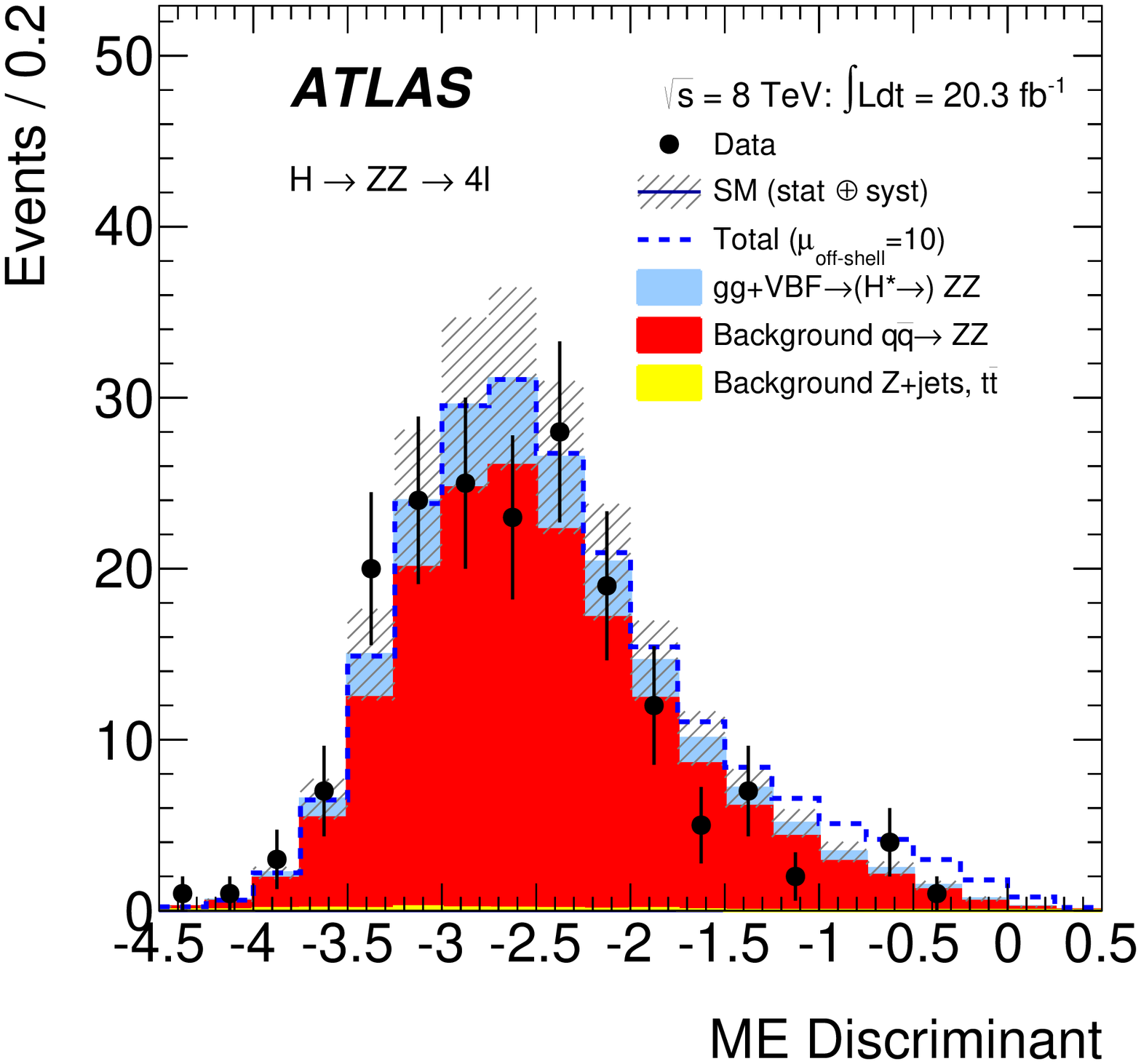}\hfil
\includegraphics[height=4.8cm, clip=true]{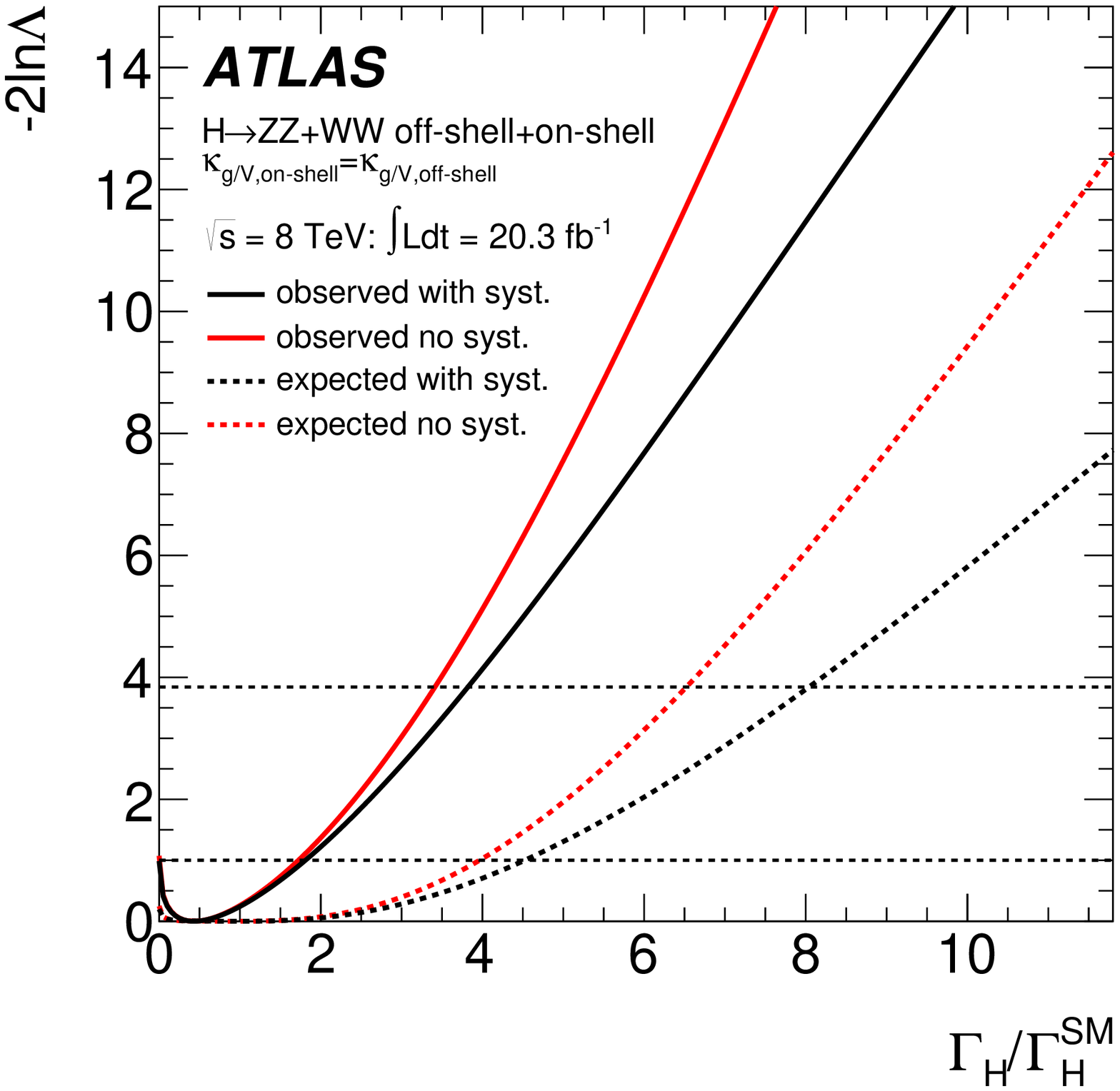}
\end{center}
\caption{ATLAS study of off-shell Higgs width constraint approach (from \cite{Aad:2015xua}).}
\label{fig:ATLAS}
\end{figure}

A proposal to exploit the Higgs width independence of the off-shell Higgs signal
in order to break the NWA scaling degeneracy
\[
\sigma_{i\to H \to f} \overset{\text{NWA}}{\propto} \frac{g_i^2 g_f^2}{\Gamma_H}\,,\quad
\text{$\sigma$ invariant if\ \ $g_i\to \xi\, g_i,\ \ g_f\to \xi\, g_f,\ \ \Gamma_H\to \xi^4\, \Gamma_H$}
\]
of the on-peak Higgs signal 
in $gg\to H\to ZZ\to 4\ell$ was first made in Ref.\ \cite{Caola:2013yja}, 
which also provided a proof-of-concept phenomenological analysis which suggested
that Higgs width constraints of $\Gamma_H < [20,38]\,\Gamma_H^\text{SM}$
are feasible.  A more detailed phenomenological analysis was subsequently
carried out in Refs.\ \cite{Campbell:2013una,Campbell:2013wga}, which 
optimized the sensitivity of the method by exploiting the full differential
cross section information using the Matrix Element Method \cite{Campbell:2012cz}
(see Fig.\ \ref{fig:MCFM}).
After these phenomenological studies, CMS \cite{Khachatryan:2014iha} (see Fig.\ \ref{fig:CMS}) and
ATLAS \cite{Aad:2015xua} (see Fig.\ \ref{fig:ATLAS}) carried out full 
experimental simulations which also took into account the $2\ell 2\nu$ final state,
vector boson fusion contributions and higher-order corrections.
CMS and ATLAS thus found 
$\Gamma_H < 5.4\,\Gamma_H^\text{SM}$ and $\Gamma_H < [4.5,7.5]\,\Gamma_H^\text{SM}$ at 95\% CL, respectively.
More recently, theorists have demonstrated that the $ZZ$+jet 
channel can be used to improve the obtained constraints \cite{Campbell:2014gua}.
That the off-shell Higgs width constraint approach is \emph{a priori} 
model dependent 
was first pointed out in Ref.\ \cite{Englert:2014aca} and has been further
studied in Refs.\ \cite{Englert:2014ffa,Logan:2014ppa}.

\begin{figure}
\begin{center}
\begin{minipage}{0.4\textwidth}\includegraphics[height=4.5cm, clip=true]{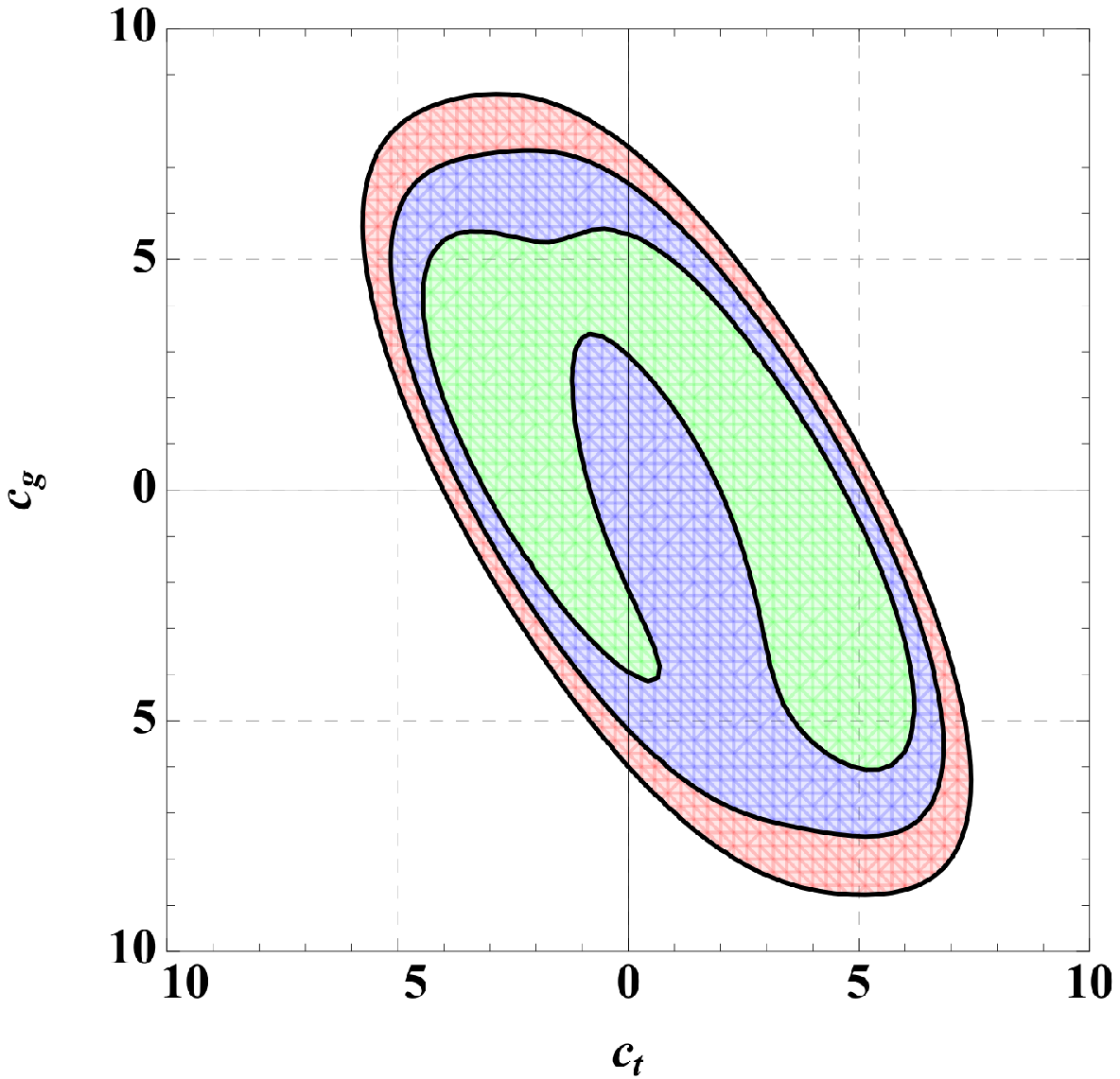}\\\centerline{\footnotesize{LHC 8 TeV CMS data}}\end{minipage}\hspace{.2cm}
\begin{minipage}{0.4\textwidth}\includegraphics[height=4.5cm, clip=true]{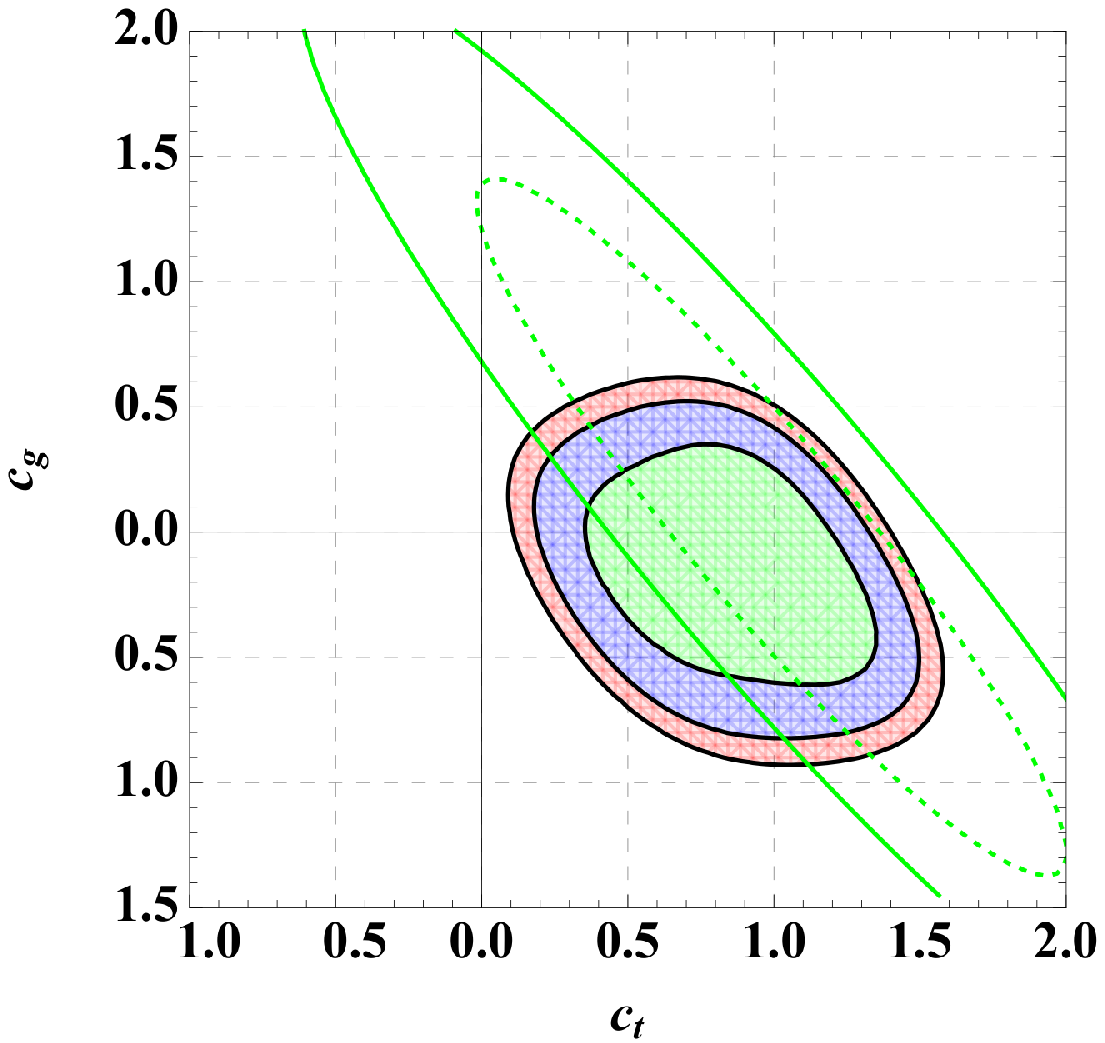}\\\centerline{\footnotesize{LHC 14 TeV\ \ 3\,ab$^{-\text{1}}$ data}}\end{minipage}
\end{center}
\caption{Constraints in $(c_t,c_g)$ plane: $68\%$, $95\%$ and $99\%$ probability contours are shown (from \cite{Azatov:2014jga}).}
\label{fig:Azatov}
\end{figure}

More generally, the off-shell $H\to VV$ signal can be used 
to disentangle degeneracies in parametric BSM studies or constrain 
higher dimensional operators in effective field theory (EFT) studies
\cite{Gainer:2014hha,Ghezzi:2014qpa,Cacciapaglia:2014rla,Azatov:2014jga,Englert:2014ffa,Biekoetter:2014jwa,Buschmann:2014sia}.
For instance, Refs.\cite{Cacciapaglia:2014rla,Azatov:2014jga,Buschmann:2014sia} analyse
SM deviations of the effective $ggH$ and $Ht\bar{t}$ coupling strengths in
an EFT approach:\\[-0.6cm]
\begin{gather*}
{\cal L}=-c_t\, \frac{m_t}{v}\,\bar t t h+\frac{g_s^2}{48 \pi^2}\, c_g\, \frac{h}{v}\,G_{\mu\nu}G^{\mu\nu}\\
\mathcal{M}_{gg\to ZZ} = \mathcal{M}_{h} + \mathcal{M}_{bkg}= c_{t}\,\mathcal{M}_{c_{t}}+c_{g}\,\mathcal{M}_{c_{g}}+\mathcal{M}_{bkg}\\[-0.8cm]
\end{gather*}
One has: $\sigma \sim |c_t+c_g|^2$. The on-peak degeneracy $c_t+c_g=\mathrm{const}$ is broken by off-shell data.  Results of Ref.\ \cite{Azatov:2014jga} are shown in 
Fig.\ \ref{fig:Azatov}.

\begin{figure}
\begin{center}
\begin{minipage}{0.42\textwidth}
\includegraphics[height=4.cm, clip=true]{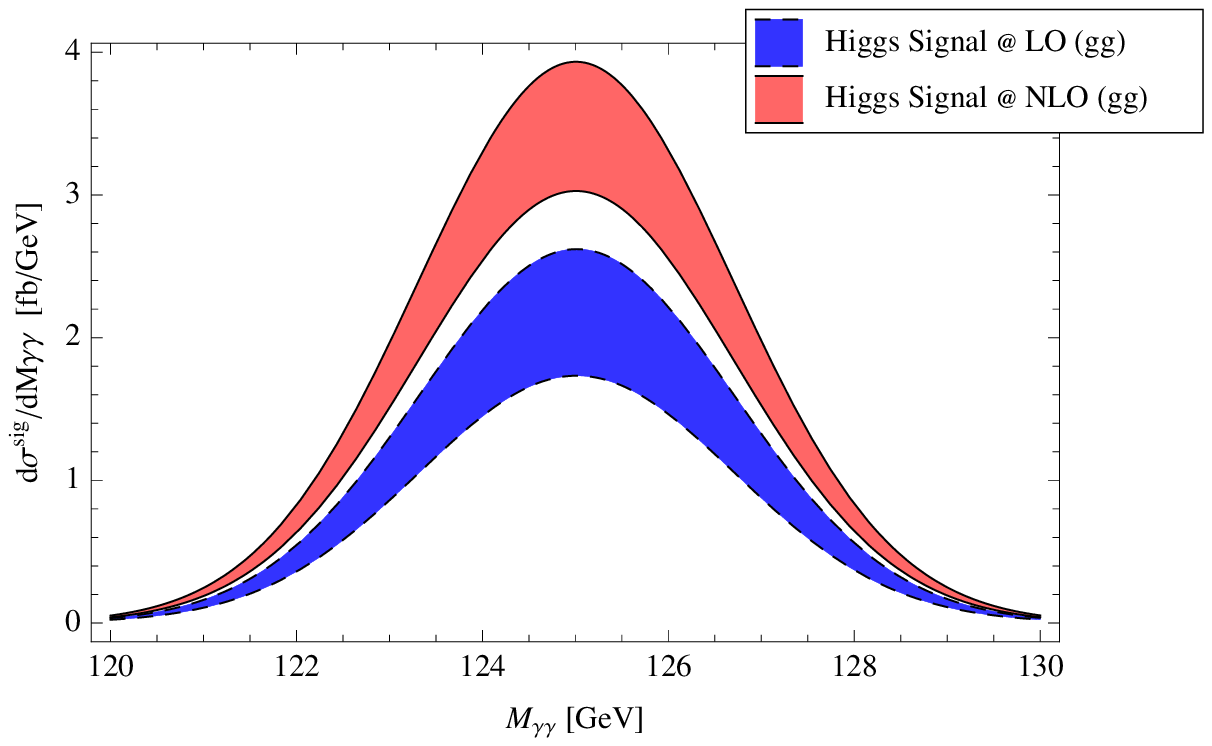}
\end{minipage}
\hfil
\begin{minipage}{0.52\textwidth}
\includegraphics[height=3cm, clip=true]{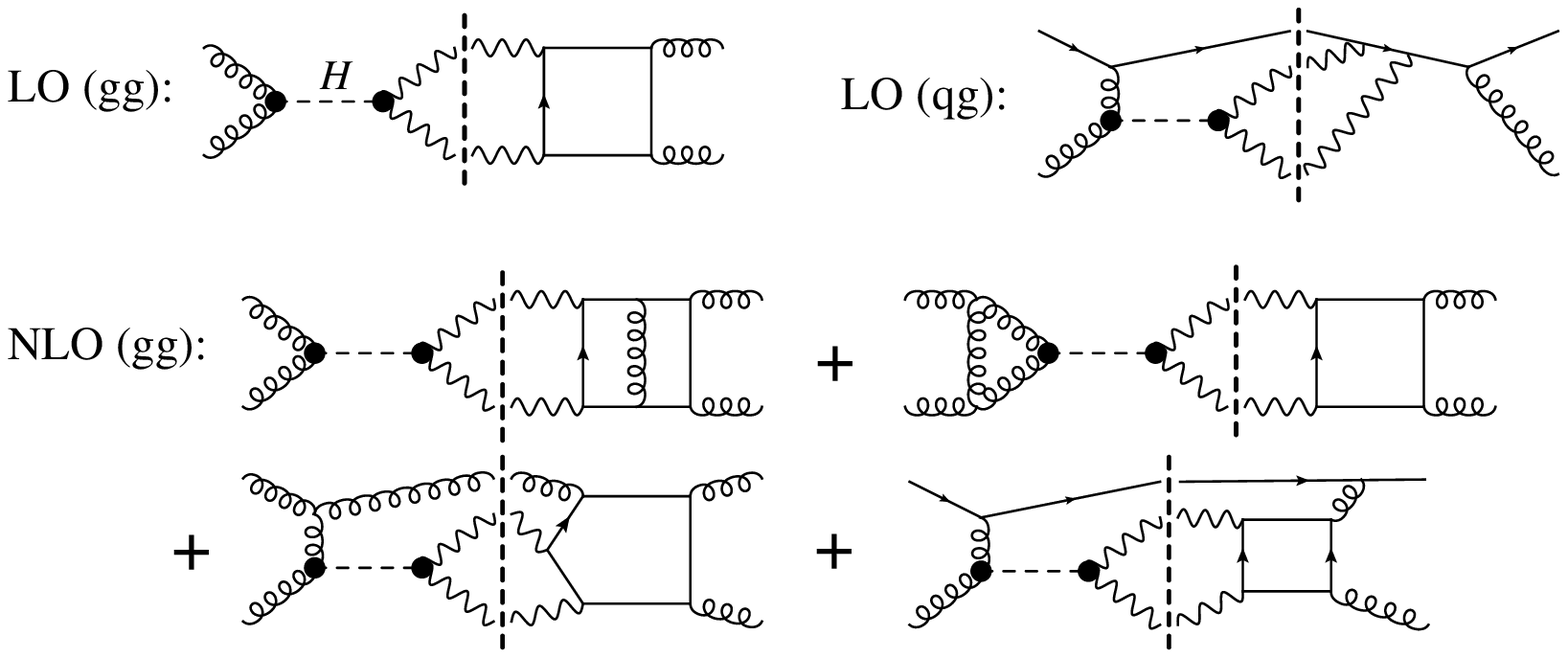}
\end{minipage}\\
\includegraphics[height=4.7cm, clip=true]{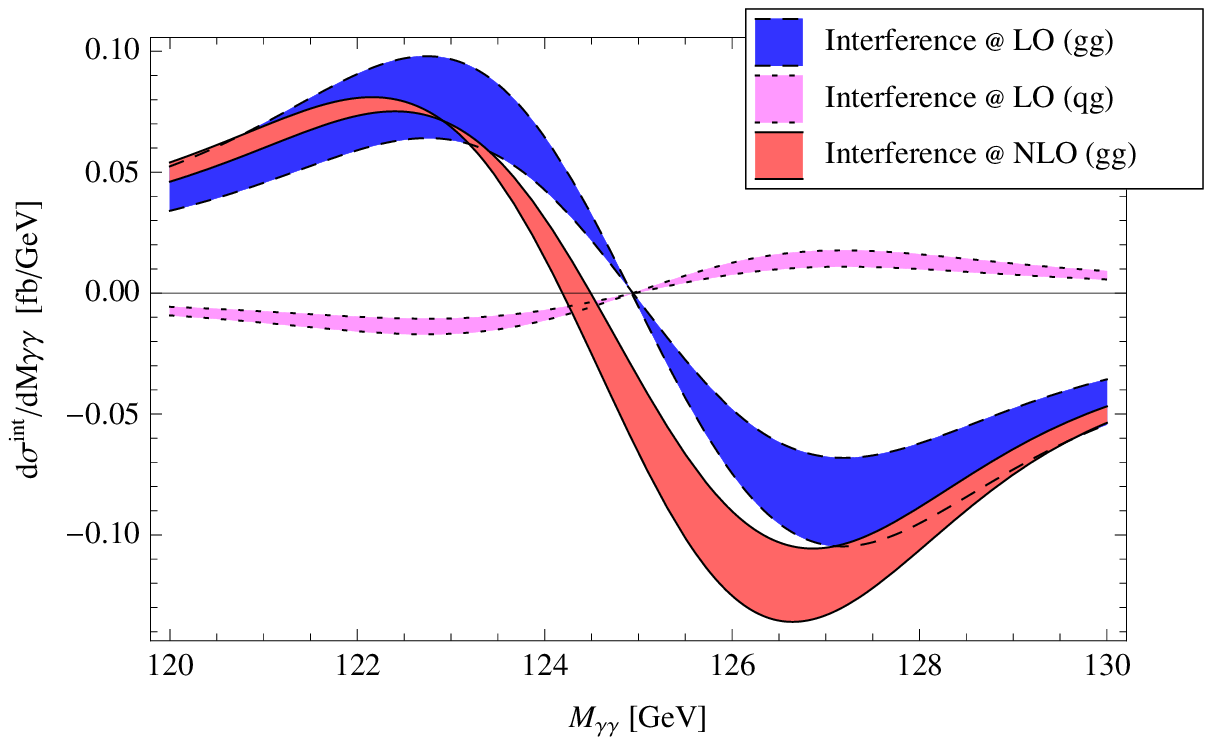}\hfil
\includegraphics[height=4.7cm, clip=true]{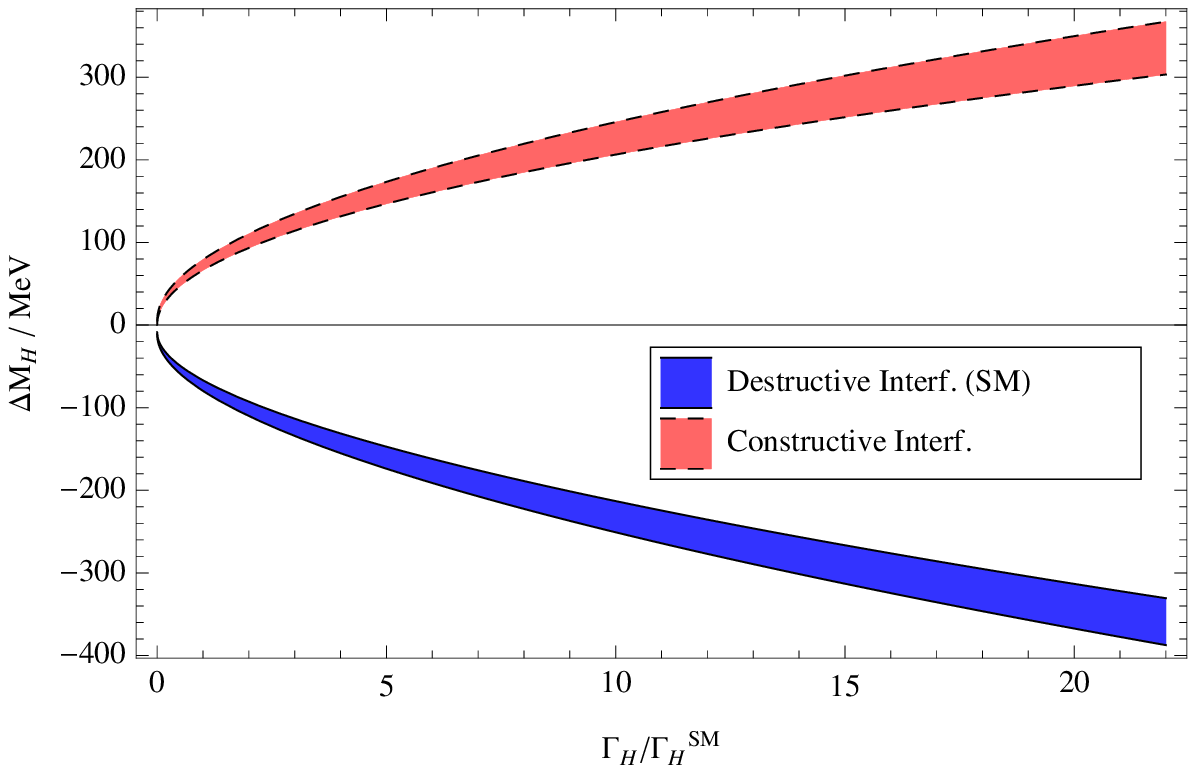}
\end{center}
\caption{Signal-background interference enabled Higgs width constraints in $gg\to H\to \gamma\gamma$ (from \cite{Dixon:2013haa})}
\label{fig:Hgammagamma}
\end{figure}

An alternative method to constrain the total Higgs width 
was proposed in Ref.\ \cite{Dixon:2013haa}.  It exploits 
a sizeable asymmetric signal-background interference in 
the $gg\to H\to\gamma\gamma$ channel at the LHC, which was 
first pointed out and calculated at LO in Ref.\ \cite{Martin:2012xc}.\footnote{%
Signal-background interference and mass peak shift effects in the $qg$ and $q\bar{q}$ channels have been analysed at LO in Refs.\ \cite{deFlorian:2013psa,Martin:2013ula}.}
A NLO calculation and analysis was carried out in 
Ref.\ \cite{Dixon:2013haa} (see Fig.\ \ref{fig:Hgammagamma}).
This method is expected to yield competitive Higgs width constraints with 
3 ab$^{-1}$ of LHC14 data.

\section{Conclusions}
Two novel, complementary methods to constrain the total Higgs width 
at the LHC have been reviewed.  The first method relies on the experimental
sensitivity to the Higgs-width-independent off-shell signal cross section
in the $gg\to H\to VV$ channels, and with LHC Run 1 data yields a Higgs width 
constraint of $\Gamma_H \lesssim 5\,\Gamma_H^\text{SM}$.  The second method relies 
on a sizeable asymmetric signal-background interference in the 
$gg\to H\to\gamma\gamma$ 
channel that results in a Higgs mass peak shift which is expected to yield 
competitive Higgs width constraints with 3 ab$^{-1}$ of LHC14 data.


\acknowledgments
Fruitful collaboration with G.\ Passarino and informative 
and useful discussions with 
A.\ Banfi, F.\ Caola, C.\ Charlot, R.\ Covarelli, S.\ Diglio, B.\ Di Micco, 
M. D\"{u}hrssen, C.\ Englert, S.\ Forte, Y.\ Gao, S.\ Gascon-Shotkin, 
D.\ Gon\c{c}alves, M.\ Grazzini, 
C.\ Grojean, S.~Heinemeyer, G.\ Isidori, S.\ J\"{a}ger, S.\ Liebler, H.\ Logan, I.\ Low, F.\ Maltoni, C.\ Mariotti, 
S.\ Marzani, M.\ M\"{u}hlleitner, B.\ Murray, C.\ O'Brien, C.\ Oleari, T.\ Plehn, M.\ Spannowsky, 
R.\ Tanaka, E.\ Vryonidou, J.\ Wang and G.\ Weiglein are gratefully acknowledged.
I would like to thank the Galileo Galilei Institute for Theoretical Physics
for hospitality and the INFN for partial support during the preparation of
this paper. This work was supported by STFC grants ST/J000485/1, ST/J005010/1 
and ST/L000512/1.


\end{document}